\documentclass[runningheads]{llncs}
\usepackage[T1]{fontenc}
\usepackage{graphicx}
\usepackage[]{hyperref} 
\usepackage{color}

\usepackage{tikz}
\usepackage{amsmath}
\usepackage{subcaption}

\usepackage{pgfplots}
\pgfplotsset{compat=1.18}
\newcommand{\qwencoderB}{blue}
\newcommand{\qwencoderF}{\qwencoderB!40}
\newcommand{\qwenB}{orange}
\newcommand{\qwenF}{\qwenB!60}
\newcommand{\llamaB}{purple}
\newcommand{\llamaF}{\llamaB!60}
\newcommand{\chatgptB}{brown}
\newcommand{\chatgptF}{\chatgptB!60}
\newcommand{\deepseekB}{red!50!black}
\newcommand{\deepseekF}{\deepseekB!60}
\newcommand{\codegptB}{olive}
\newcommand{\codegptF}{\codegptB!60}
\newcommand{\codegenB}{green!40!black}
\newcommand{\codegenF}{\codegenB!50}
\newcommand{\codetfiveB}{magenta}
\newcommand{\codetfiveF}{\codetfiveB!60}

\usepackage{tabularx}
\usepackage{makecell}
\usepackage{booktabs}
\usepackage{ragged2e}
\renewcommand{\arraystretch}{1.3}
\renewcommand{\theadfont}{\footnotesize\bfseries}
\usepackage{siunitx} 
\usepackage{pifont}
\definecolor{darkgreen}{RGB}{0,150,0}
\definecolor{darkred}{RGB}{230, 0, 0}

\usepackage{xspace}
\usepackage{enumitem}
\begin{document}
\date{}
%
\title{Towards Automated Pentesting with Large Language Models}
\author{Ricardo Bessa\inst{1} \and Rui Claro\inst{2} \and João Trindade\inst{2} \and João M. Lourenço\inst{1}}
\institute{NOVA University Lisbon — FCT \& NOVA LINCS, Portugal \and Layer8 - Shield Domain SA, Portugal \\[1mm]
\email{r.bessa@campus.fct.unl.pt, joao.lourenco@fct.unl.pt,\\\{rui.claro,joao.trindade\}@layer8.pt}}
\authorrunning{R. Bessa et al.}
\maketitle             
\begin{abstract}
Large Language Models (LLMs) are redefining offensive cybersecurity by allowing the generation of harmful machine code with minimal human intervention. While attackers take advantage of dark LLMs such as XXXGPT and WolfGPT to produce malicious code, ethical hackers can follow similar approaches to automate traditional pentesting workflows. In this work, we present RedShell, a privacy-preserving, hardware-efficient framework that leverages fine-tuned LLMs to assist pentesters in generating offensive PowerShell code targeting Microsoft Windows vulnerabilities. RedShell was trained on a malicious PowerShell dataset from the literature, which we further enhanced with manually curated code samples. Experiments show that our framework achieves over 90\% syntactic validity in generated samples and strong semantic alignment with reference pentesting snippets, outperforming state-of-the-art counterparts in distance metrics such as edit distance (above 50\% average code similarity). Additionally, functional experiments emphasize the execution reliability of the snippets produced by RedShell in a testing scenario that mirrors real-world settings. This work sheds light on the state-of-the-art research in the field of Generative AI applied to malicious code generation and automated testing, acknowledging the potential benefits that LLMs hold within controlled environments such as pentesting.

\end{abstract}
%
\section{Introduction}

In recent years, LLMs sparked a revolution in natural language processing through advanced transformers~\cite{Vaswani2017}. Models such as ChatGPT~\cite{chatgpt} and DeepSeek~\cite{deepseek} have demonstrated strong capabilities while performing generic tasks, including text classification and language understanding~\cite{Wolf2020}. In addition, fine-tuning techniques have been employed to adapt LLMs to address more specific challenges, with Generative AI becoming pervasive in many software-engineering workflows, particularly in automated testing~\cite{Fan2023}.

In the offensive cybersecurity field, AI models such as XXXGPT and WolfGPT arise as a new tool for malicious actors to generate harmful code~\cite{Rustam2024}. However, ethical hackers can also take advantage of LLMs to develop malicious code generators, providing more automation to traditional pentest audits. Penetration testing, often referred to as pentesting, is a crucial activity for red teams, which are responsible for testing the cybersecurity effectiveness of a system through simulated cyber-attacks~\cite{Vats2020}. By detecting potential security lapses in a target system, pentesters are able to prevent attacks that could cause harm to that system and its users. Ethical hackers often rely on Assembly shellcode, PowerShell scripts, and other forms of malicious code to identify and exploit security vulnerabilities. However, regardless of the exploitation technique being employed, writing offensive code demands significant time, effort, and expertise from pentesters~\cite{Nautilus}. While LLMs offer a promising path towards automated malicious code generation, this approach also remains challenging due to the limited public availability of offensive training data.

In this paper, we present RedShell, a framework that incorporates locally fine-tuned LLMs to support malicious PowerShell code generation, providing more automation to pentesting activities targeting Microsoft Windows. To build RedShell, we followed a privacy-preserving and hardware-efficient methodology that can be generalized to support the production of other forms of offensive code beyond PowerShell. We also build on a malicious PowerShell dataset from the literature, extending it with code samples from various cybersecurity frameworks, to train and evaluate models in offensive PowerShell code generation.

The evaluation focuses on assessing the quality of the generated snippets by examining their syntactic, semantic, and functional correctness, as well as performing a comparative analysis of the generation capabilities of different LLMs. Experiments show that our framework was able to generate syntactically valid PowerShell code, with over 90\% of the generated samples successfully parsed without errors. Additionally, RedShell produced samples closely aligned with reference snippets, achieving more than 50\% average code similarity across standard metrics such as edit distance. These results position RedShell as a competitive solution for automated malicious PowerShell generation, outperforming the output semantics of reference models in the cybersecurity domain. Furthermore, functional experiments conducted within a controlled environment demonstrate the execution reliability of the snippets produced by RedShell in a realistic pentesting scenario.

The remainder of this paper is organized as follows: Section~\ref{sec:related_work} reviews background concepts and related work; Section~\ref{sec:ground_truth_dataset} introduces an extended version of a malicious PowerShell dataset from the literature; Section~\ref{sec:redshell_design} details the design of RedShell; Section~\ref{sec:evaluation} presents experiments validating RedShell’s generation capabilities; Section~\ref{sec:ethical_considerations} discusses ethical considerations regarding malicious AI usage, and finally, Section~\ref{sec:conclusions} concludes the paper and describes future directions.
%
\section{Background \& Related Work}
\label{sec:related_work}

LLMs registered a high popularity in recent years, with the AI-community actively searching for new approaches to develop generative models that deliver strong performance on both general and domain-specific tasks. In addition, engineers are taking advantage of these novel technologies to automate complex and time-consuming processes, allowing human operators to focus on more critical aspects of their workflows. We now discuss core concepts within the fields of pentesting and Generative AI, and then we review related work on the application of LLMs in cybersecurity.

\subsection{Traditional Pentesting}

\begin{description}[style=unboxed,leftmargin=0cm,itemsep=7pt]

  \item[Pentesting.] While conducting pentest audits, ethical hackers follow an offensive methodology that starts from analyzing a target network or system, and culminates in the emulation of real adversary behaviors~\cite{Vats2020}. Through pentesting, red teams can assess the functional aspects of a system and detect potential security vulnerabilities that could be exploited by real attackers. Weidman~\cite{weidman} highlights the differences between pentesting and vulnerability assessment. In a vulnerability assessment, security professionals identify core system vulnerabilities that could be exploited by malicious actors. Pentesting extends vulnerability assessment by further performing the exploitation of the detected flaws to assess what hackers might achieve after a successful attack.

  \item[MITRE ATT\&CK.] The MITRE ATT\&CK framework~\cite{mitre_attack} provides a comprehensive taxonomy of the tactics, techniques, and procedures (TTPs) employed by real adversaries to conduct offensive operations. The framework classifies malicious activities into 14 tactical groups representing different stages of an attack. Built on real-world observations, MITRE ATT\&CK offers a valuable knowledge base for security professionals to develop threat models in both the private sector and the cybersecurity product and service community~\cite{Strom2018}. The framework covers the adversarial behaviors that can be performed against different operating systems, including detailed information describing the most common tools and methodologies to carry out those attacks.

  \item[PowerShell.] As Microsoft Windows remains one of the most frequently targeted operating systems, PowerShell has evolved into an effective exploitation framework for pentesters~\cite{Liguori2024}. The PowerShell programming language is included by default in modern versions of Windows, providing a flexible scripting environment for automating administrative tasks and managing remote services. PowerShell's versatility also makes it an attractive vector for adversaries, who can leverage its built-in functionality to execute custom payloads without downloading additional malicious tools on the target system. In 2024, PowerShell ranked fourth among the most frequently observed MITRE ATT\&CK techniques in confirmed threats, affecting 20.4\% of customers and accounting for 684 detected incidents across the Red Canary customer base~\cite{redcanary2024}. Reported occurrences showcase malicious agents actively abusing PowerShell to obfuscate offensive activities, gather information, change system configurations, and remotely download and execute arbitrary code and binaries such as Mimikatz~\cite{Mimikatz} (to dump in-memory secrets), and  AzureAD~\cite{AzureAD} (to target cloud services). Furthermore, since Windows 10 reached its end of support in 2025, unpatched systems are likely to increase, creating a larger attack surface for PowerShell-based exploitation.

\end{description}

\subsection{Generative AI}

\begin{description}[style=unboxed,leftmargin=0cm,itemsep=7pt]

  \item[Pattern Recognition.] Early pattern recognition techniques laid the foundation for Generative AI. Pattern recognition consists of the automatic discovery of regularities in data through the use of computer algorithms. Detected similarities can be used to take actions and perform tasks with more automation. Throughout the years, many strategies were proposed to deal with the problem of searching for patterns in data. However, the best results were achieved by adopting a Machine Learning approach~\cite{bishop}.

  \item[Machine Learning (ML).] In ML, a large set of $n$ data points $\{x_1, \ldots, x_n\}$, known as the training dataset, is used to tune the parameters of an adaptive model. The categories of those data points are known in advance, typically by individual inspection and manual labeling. The training dataset can be used to build a ML algorithm, which can be expressed as a function $y(x)$ that takes a new data input $x$ and generates an output vector $y$. The precise form of the function $y(x)$ is determined during the learning phase, on the basis of the training data. An epoch is a complete period of training the model with the training dataset. Multiple epochs allow the model to revisit the same data and get more accurate learning. The trained model can then be used to predict the label of new $x$ inputs, which are part of the test dataset. The capability of the model to produce correct predictions is known as generalization.

  \item[Transformers.] While Deep Learning emerged in the 1980s as a novel approach for computers to learn complex concepts from simple representations, proposed models were limited by the computing power available at the time~\cite{Goodfellow}. However, recent advances in computational infrastructure have enabled the training of larger neural networks on massive text corpora. Furthermore, transformer architectures~\cite{Vaswani2017} introduced an attention mechanism connecting encoder and decoder layers of Deep Learning models, overcoming the scalability bottleneck of traditional recurrent networks. Today, transformers became the dominant architecture to build LLMs.

  \item[Neural Machine Translation (NMT).] The field of Natural Language Processing (NLP) aims to convert human language into a formal representation that is easy for computers to manipulate~\cite{Collobert2008}. Taking advantage of NLP techniques, LLMs are able to process user prompts and produce original responses, which is the basis to achieve Generative AI\@. NLP is composed of different sub-tasks, including NMT, which consists of the translation of a sequence of tokens $x$ in a given source language, into a sequence of tokens $y$ in a target language~\cite{Bahdanau2015}. The translation is performed for every individual token by choosing the prediction with the highest probability from a set of translation probabilities $P(y|x)$. NMT models are trained to predict the next word $y_{t'}$ given the context vector $c$ and all the previously predicted words $\{y_1, \cdots, y_{t'-1}\}$. Assuming $y = (y_1, \cdots, y_{T_y})$, the probability over the translation $y$ is defined by decomposing the joint probability into the ordered conditionals $p(y) = \prod_{t=1}^{T} p(y_t \mid \{y_1, \cdots, y_{t-1}\}, \;c)$.

\end{description}

\subsection{Generative AI and Cybersecurity}

\begin{description}[style=unboxed,leftmargin=0cm,itemsep=7pt]

  \item[Defensive Cybersecurity.] Blue teams can integrate LLMs into their workflows to streamline security risk assessments and generate real-time responses to cyber-threats~\cite{Nourmohammadzadeh2024}. By processing large volumes of data, LLMs can deliver accurate and predictive risk assessments, detecting subtle patterns that human analysts might overlook. For instance, Eze and Shamir~\cite{Eze2024} leverage AI models to identify AI-generated phishing emails through automated text analysis, while Hanif and Maffeis~\cite{Hanif2022} propose VulBERTa, a model specialized in binary and multi-class vulnerability detection in C$/$C$++$ code. Moreover, Sladić \textit{et al.}~\cite{Sladic2023} introduce shellLM, a LLM-based solution for generating Linux shell honeypots. Defensive cybersecurity approaches fall outside the scope of this work. \\[-1.5ex]

  \item[Offensive Cybersecurity.] Red teams can leverage LLM agents such as PentestGPT~\cite{Deng2023} and PENTEST-AI~\cite{Bianou2024} to automate complex pentest audits. Agentic frameworks are typically composed of a set of LLM-based modules that perform specific tasks within the pentesting process while working together to conduct a complete and effective attack. In particular, such frameworks offer mechanisms to detect, explore, and report different vulnerabilities with minimal human intervention. RedShell adopts a more fine-grained approach, offering a specialized solution to assist ethical hackers in malicious PowerShell generation, and potentially serving as a building block for broader autonomous pentesting agents. Some literature also focused on the application of LLMs in offensive code generation. EVIL~\cite{Liguori2021} fine-tuned NMT models in malicious Assembly generation, using an extended collection of Assembly shellcodes and Python encoders built on top of the Shellcode\_IA32 dataset~\cite{Liguori2022}. The generation of shellcodes was also targeted by tools such as DualSC~\cite{Yang2022} and ExploitGen~\cite{Yang2023}, using a shallow transformer and a template-augmented approach, respectively. In contrast to DualSC, which introduces a custom transformer, RedShell leverages existing state-of-the-art LLM architectures. Additionally, unlike ExploitGen, our framework does not rely on a template parser, thus avoiding the associated parsing overhead and enabling faster model inference. Natella~\textit{et al}.~\cite{Natella2024} introduced a manually curated dataset to train models in offensive Python generation. Chowdhary, Jha, and Zhao~\cite{Chowdhary2023} employed Generative Adversarial Networks (GANs) to automate XSS attacks on web applications, while Liguori \textit{et al.}~\cite{Liguori2024} combined pre-training and fine-tuning strategies to generate malicious PowerShell snippets using LLMs. However, both approaches rely on computationally intensive training strategies. The use of GANs by Chowdhary \textit{et al.} in~\cite{Chowdhary2023} requires training multiple competing neural networks, and the work by Liguori~\cite{Liguori2024} involves pre-training models on a large generic PowerShell corpus. Pre-training LLMs is known to be highly resource-demanding. For instance, Khandelwal \textit{et al.}~\cite{khandelwal} report that pre-training Pythia-1B required 64 GPUs for three days, while RoBERTa required \num{1000} GPUs for one day. In contrast, RedShell adopts a lightweight, hardware-efficient training methodology based solely on fine-tuning techniques with a single GPU.

\end{description}

\section{Ground Truth Dataset}
\label{sec:ground_truth_dataset}

\begin{table}[t]
    \caption{GitHub popularity of PowerShell, Java, and Python (November, 2025).}
    \vspace{1mm}
    \label{tab:github}
    \centering
    {\footnotesize
        \renewcommand{\arraystretch}{1.3}
        \renewcommand{\theadfont}{\footnotesize\bfseries}
        \hyphenpenalty=10000
         \begin{tabular}{
            l
            S
            S
            S
        }
        \toprule
        \thead[l]{~~~Language} & {\thead{~~~~Repos (k)}~~~~~~~} & {\thead{Wikis (k)}~~~~~} & {\thead{Users (k)~~~~}} \\
        \midrule
                ~~~PowerShell  &   63    & 33    & 1     \\
                ~~~Java        &  2800    & 340   & 74    \\
                ~~~Python      &  3500    & 401   & 104   \\
        \bottomrule
        \end{tabular}
    }
\end{table}

We now present both reference and extended versions of a ground truth dataset to support the training and evaluation of RedShell.

\subsection{Reference Dataset}

To train and evaluate RedShell, we developed a ground truth dataset pairing malicious PowerShell snippets with natural language descriptions. These descriptions are designed to capture the semantic intent of each snippet, specifically identifying the offensive techniques used and the vulnerabilities targeted. Constructing this dataset from scratch presented a significant challenge due to the scarcity of PowerShell code repositories compared to ubiquitous languages like Java or Python. As illustrated in Table~\ref{tab:github}, there is a stark disparity in the availability of resources between the general-purpose languages and the domain-specific PowerShell environment.

Based on the previous observation, we selected~\cite{dessertlab} as a reference corpus to support the training and evaluation of the models incorporated in our framework. The dataset is composed of \num{1127} samples of offensive PowerShell commands collected from online wikis and community blogs about ethical hacking such as HackTricks~\cite{hacktricks} and Infosec Matter~\cite{infosecmatter}, and from various offensive cybersecurity frameworks, including Atomic Red Team~\cite{AtomicRedTeam} and Empire~\cite{empire_project}. Each PowerShell snippet is annotated with a natural language (English) description extracted from the respective code source, while additional information was provided to the samples that did not come with a predefined label or lacked more clear descriptions.

The reference dataset was employed by Liguori \textit{et al.}~\cite{Liguori2024} to effectively train LLMs in performing NMT from natural language descriptions into malicious PowerShell snippets, addressing a previously unexplored research gap. However, we identified the following limitations and improvement opportunities regarding the contents of the reference corpus.

\begin{description}[style=unboxed,leftmargin=0cm,itemsep=7pt]

  \item[Size Limitation.] The dataset provides a limited number and diversity of code samples, which may limit the efficacy of the fine-tuning process.

  \item[Module Coverage.] Relevant offensive PowerShell modules were not covered by the dataset, including Nishang~\cite{nishang}, PowerUpSQL~\cite{powerupsql}, and MicroBurst~\cite{MicroBurst}.

  \item[MITRE ATT\&CK Coverage.] The \textit{Resource Development} tactic was not covered by the dataset. Moreover, a limited number of snippets were provided for relevant pentesting tactics, including \textit{Execution} (54 snippets), \textit{Privilege Escalation} (37 snippets), and \textit{Reconnaissance} (5 snippets).

\end{description}

\begin{table}[t]
    \centering
    \caption{Illustrative snippets from the extended ground truth dataset.}
    \vspace{1mm}
    \label{tab:snippets_ground_truth}
	\begin{tabularx}{\columnwidth}{>{\scriptsize}l   >{\RaggedRight\itshape\scriptsize}X   >{\RaggedRight\ttfamily\scriptsize}X}
		\toprule
	    	\thead[l]{Tactic} & \thead{\textup{Code Description}} & \thead{Code Snippet} \\
        \midrule
	    	\makecell[tl]{Discovery} & List the members of the Admins domain group. & Get-NetGroupMember -GroupName "Admins"; Get-ADGroupMember -Identity "Admins"\\
	    \midrule
	   	       \makecell[tl]{Defense\\Evasion} & Encode a ps command in Base64. & \$cmd=[Convert]::ToBase64String([System .Text.Encoding]::Unicode.GetBytes("\$ps")) \\
	    \midrule
	    	\makecell[tl]{Credential\\Access} & Use Mimikatz to dump secrets from the Security Account Manager. & Invoke-Mimikatz -Command "lsadump::sam"\\
        \midrule
            \makecell[tl]{Execution} & Download script.ps1 from a webserver into RAM. & iex (New-Object~Net.WebClient) .DownloadString("http://ip/script.ps1")\\
		\bottomrule
	\end{tabularx}
\end{table}

\subsection{Extended Dataset}

We built an extended version of the reference ground truth corpus by collecting new malicious samples, thus improving the overall quality and size of the dataset. We introduced \num{1135} additional code samples of offensive PowerShell, effectively doubling the size of the original corpus. Table~\ref{sec:ground_truth_dataset} presents illustrative snippets from the extended dataset. The newly collected code samples provide a broader view on existing offensive PowerShell modules, including Nishang~\cite{nishang}, PowerUpSQL~\cite{powerupsql}, and MicroBurst~\cite{MicroBurst}. Furthermore, additional snippets were extracted from tryhackme~\cite{TryHackMe} walkthroughs on capture the flag (CTF) challenges, and from GitHub repositories containing malicious PowerShell.

\begin{figure}[h]
    \begin{tikzpicture}
        \begin{axis}[
            ybar stacked,
            bar width=10pt,
            legend style={
                at={(0.95,0.95)},
                anchor=north east,
                font=\scriptsize,
                /tikz/every node/.style={anchor=west}
            },
            symbolic x coords={
                Discovery, Defense Evasion, Credential Access, Execution,
                Lateral Movement, Reconnaissance, Command and Control,
                Persistence, Privilege Escalation, Collection, Impact,
                Exfiltration, Resource Development, Initial Access
            },
            xtick=data,
            x tick label style={
                rotate=45,
                anchor=east,
                font=\scriptsize,
                yshift=-5pt
            },
            ymin=0, ymax=750,
            ytick={0,100,200,300,400,500,600,700},
            ylabel={Number of Samples},
            width=\columnwidth,
            height=0.45\columnwidth,
            enlarge x limits=0.05,
            xtick style={draw=none},
            ytick style={draw=none}
        ]

            \addplot[fill=gray] coordinates {
                (Discovery, 205)
                (Defense Evasion, 430)
                (Credential Access, 163)
                (Execution, 54)
                (Lateral Movement, 96)
                (Reconnaissance, 5)
                (Command and Control, 38)
                (Persistence, 42)
                (Privilege Escalation, 37)
                (Collection, 32)
                (Impact, 12)
                (Exfiltration, 10)
                (Resource Development, 0)
                (Initial Access, 4)
            };

            \addplot[fill=gray!30] coordinates {
                (Discovery, 476)
                (Defense Evasion, 151)
                (Credential Access, 109)
                (Execution, 105)
                (Lateral Movement, 22)
                (Reconnaissance, 77)
                (Command and Control, 36)
                (Persistence, 31)
                (Privilege Escalation, 35)
                (Collection, 33)
                (Impact, 19)
                (Exfiltration, 15)
                (Resource Development, 18)
                (Initial Access, 8)
            };

            \node at (axis cs:Discovery,681) [anchor=south, font=\scriptsize] {681};
            \node at (axis cs:Defense Evasion,581) [anchor=south, font=\scriptsize] {581};
            \node at (axis cs:Credential Access,272) [anchor=south, font=\scriptsize] {272};
            \node at (axis cs:Execution,159) [anchor=south, font=\scriptsize] {159};
            \node at (axis cs:Lateral Movement,118) [anchor=south, font=\scriptsize] {118};
            \node at (axis cs:Reconnaissance,82) [anchor=south, font=\scriptsize] {82};
            \node at (axis cs:Command and Control,74) [anchor=south, font=\scriptsize] {74};
            \node at (axis cs:Persistence,73) [anchor=south, font=\scriptsize] {73};
            \node at (axis cs:Privilege Escalation,72) [anchor=south, font=\scriptsize] {72};
            \node at (axis cs:Collection,65) [anchor=south, font=\scriptsize] {65};
            \node at (axis cs:Impact,31) [anchor=south, font=\scriptsize] {31};
            \node at (axis cs:Exfiltration,25) [anchor=south, font=\scriptsize] {25};
            \node at (axis cs:Resource Development,18) [anchor=south, font=\scriptsize] {18};
            \node at (axis cs:Initial Access,12) [anchor=south, font=\scriptsize] {12};

            \legend{Reference Samples~\cite{Liguori2024}, New Samples}
        \end{axis}
    \end{tikzpicture}
    \caption{Dataset coverage of the MITRE ATT\&CK tactics.}
    \label{fig:mitre-mapping}
\end{figure}
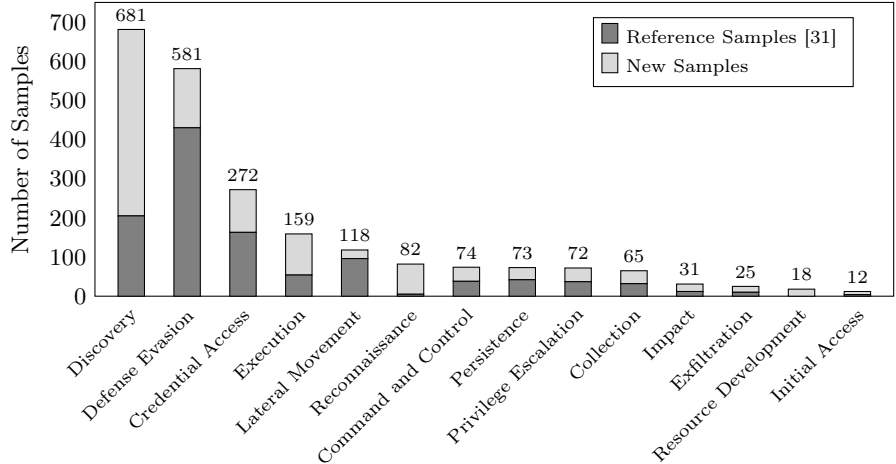

We manually classified the tactical group of each new code snippet following the MITRE ATT\&CK documentation. For the reference samples, we adopted the classifications provided by Liguori~\textit{et al.} in~\cite{Liguori2024}. Figure~\ref{fig:mitre-mapping} reports the number of PowerShell samples per offensive tactic, highlighting how both reference and extended datasets contribute to the final ground truth corpus. The extended dataset not only covers all the~14 offensive tactics from the MITRE ATT\&CK framework, but also increments the number of available samples for the offensive techniques typically employed by pentesters. In particular, the tactics that registered the largest increases in sample count were \textit{Discovery} (476$+$ snippets), \textit{Defense Evasion} (151$+$ snippets), and \textit{Credential Access} (109$+$ snippets). Relevant pentesting tactics such as \textit{Command and Control}, \textit{Persistence}, and \textit{Privilege Escalation} also experienced significant improvements, with their sample counts nearly doubling. 

We now examine the most prevalent tactics in the dataset, highlighting the PowerShell utilities that pentesters may leverage in post-exploitation scenarios.

\begin{description}[style=unboxed,leftmargin=0cm,itemsep=7pt]

  \item[Discovery.] Represents almost half of the number of new collected snippets. The high availability of \textit{Discovery} samples highlights the capabilities offered by PowerShell to perform various discovery activities on Microsoft Windows, including network service discovery and permission groups discovery.

  \item[Defense Evasion.] The dataset offers a wide representation of defensive strategies employed by ethical hackers to avoid detection and conceal traces of their malicious activities. \textit{Defense Evasion} techniques include disabling or modifying security software such as the Windows Defender and the Antimalware Scan Interface (AMSI), to allow the stealthy execution of malicious programs. Pentesters might also create Firewall rules to allow incoming malicious traffic into the target network or try to obfuscate malicious files to bypass security detectors.

  \item[Credential Access.] Security professionals take advantage of \textit{Credential Access} techniques to dump credentials and obtain account names and passwords. Legitimate credentials can then be used to access sensitive information and services while making malicious activities harder to detect.

  \item[Execution.] Ethical hackers typically employ \textit{Execution} techniques to run malicious code on a local or remote system. For instance, pentesters can take advantage of PowerShell utilities to download and run malicious scripts in-memory, creating a fileless malware that is harder to detect since it does not leave persistent traces on disk.

\end{description}

The ground truth dataset was randomly split in two partitions, the training and the test datasets. Since the fine-tuning was conducted with a small and manually curated corpus, we adopted a 90/10 train-test split to maximize the amount of data available for training, optimizing the intended knowledge transfer. The test partition, while comprising a significantly smaller part of the dataset, still provides a sufficient amount of unseen data to assess the generalization capabilities of the fine-tuned models and detect potential overfitting scenarios. Additionally, we note that the train-test split of the dataset does not guarantee balanced representation across MITRE ATT\&CK tactics. However, since the most common pentesting tactics such as \textit{Discovery}, \textit{Defense Evasion}, and \textit{Credential Access} dominate the dataset, random partitioning still ensures a high likelihood of their presence on both partitions.

\section{RedShell Design}
\label{sec:redshell_design}

\begin{figure}[t]
    \centering
    \includegraphics[scale=0.07]{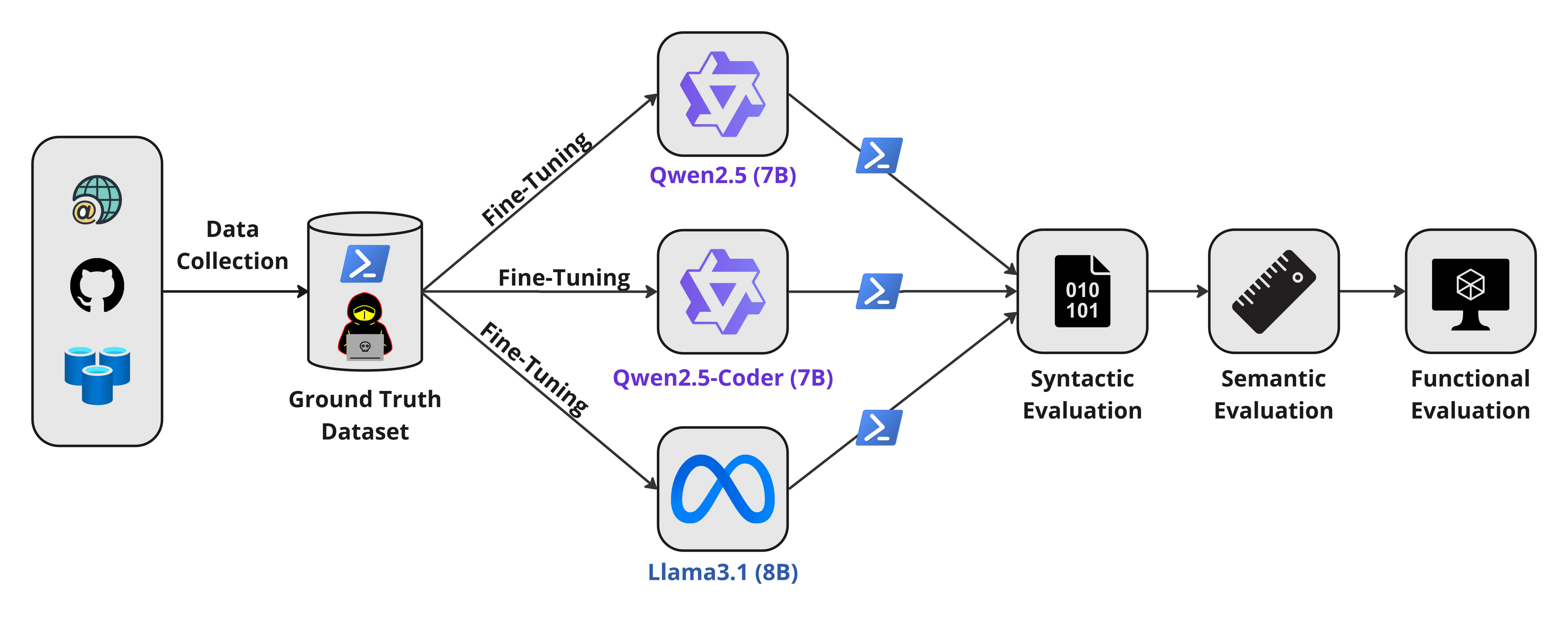}
    \caption{Overview of the RedShell framework architecture.}
    \label{fig:design}
\end{figure}

Previously, we introduced an offensive PowerShell ground truth dataset, a crucial component for building a malicious code generator such as RedShell. We now introduce the remaining modules of our framework (Figure \ref{fig:design}), including the process for fine-tuning a set of selected LLMs in malicious PowerShell generation, and their performance evaluation.

\subsection{Models}

We selected three different models to be fine-tuned in malicious PowerShell generation, using both the reference and extended versions of the ground truth dataset. The selection process was performed based on the following criteria.

\begin{description}[style=unboxed,leftmargin=0cm,itemsep=7pt]
  \item[State-of-the-art.] We prioritized models representing the state-of-the-art families and architectures of LLMs, also considering their release date and popularity within the AI research community. In particular, we selected models that ranked high on the Hugging Face Open LLM Leaderboard~\cite{leaderboard} and were among the most downloaded during the first half of 2025. Only models released between the second half of 2024 and the first half of 2025 were considered.

  \item[Availability.] To support a local fine-tuning approach that ensures privacy of all training data, we only considered open-weights models i.e., LLMs with publicly available weights.

  \item[Coding Skills.] We selected models already pre-trained in multiple programming languages to perform general-purpose coding tasks, providing a strong base knowledge in code generation, summarization and reasoning.

  \item[Model Size.] We chose models with relatively small sizes, offering the possibility of being fine-tuned with less computational resources. Specifically, we considered LLMs with a parameter count up to 8 billions.
\end{description}

Given the above criteria, we selected the following LLMs to be fine-tuned in malicious PowerShell generation.

\begin{description}[style=unboxed,leftmargin=0cm,itemsep=7pt]
  \item[Qwen2.5-7B~\cite{qwen25}] belongs to the general-purpose series of Qwen LLMs, offering significantly more knowledge than the previous versions while providing improved capabilities in coding and mathematics.

  \item[Qwen2.5-Coder-7B-Instruct~\cite{qwenCoder}] belongs to the code-spe\-cif\-ic series of Qwen LLMs, offering strong capabilities in code generation, reasoning and fixing.

  \item[Llama3.1-8B~\cite{llama}] was released on July of 2024 as part of the multilingual collection of LLMs from Meta, outperforming many of the available open-source and closed models on common industry coding benchmarks.
\end{description}

The selected models were downloaded from Hugging Face and fine-tuned locally, following a training strategy that overcomes the limitations presented by closed models such as ChatGPT. When compared to proprietary LLMs, our framework offers the following properties.

\begin{description}[style=unboxed,leftmargin=0cm,itemsep=7pt]

  \item[Data Privacy.] Closed models fail to provide data privacy guarantees since any prompts submitted during pentesting activities may reveal confidential information regarding audit targets, including hostnames and IP addresses. In contrast, RedShell adopts a privacy-preserving setup where both fine-tuning and inference processes are exclusively executed on local machines, ensuring that no sensitive data leaves the testing environment.

  \item[Specialization.] LLMs trained on large-scale text corpora achieve a strong performance on more generic tasks, while RedShell was specifically designed to assist pentesters in malicious PowerShell code generation, providing a highly specialized solution that is aligned with the most common pentesting strategies.

  \item[Weak Ethical Boundaries.] Proprietary models are usually protected by ethical boundaries that restrict their generation capabilities. Although these protections can be often bypassed through prompt-engineering techniques, that manipulation requires time and effort from the pentesters. In contrast, the behavior of our specialized models was specifically adapted to support the automation of pentesting activities. The potential risks associated with the unethical use of our framework are further discussed in Section~\ref{sec:ethical_considerations}.

  \item[Generic Architecture.] RedShell is built from a set of modular, reusable components that can be easily adapted to generate other forms of malicious code beyond PowerShell while following the same design pipeline (Figure~\ref{fig:design}). For instance, extending our framework to produce malicious Assembly requires only replacing the training dataset with a custom one, the syntax checker with an Assembly validator, and modifying the functional testing environment to support Assembly code execution. All intermediate steps, namely the training process and semantic evaluation, remain unchanged.
\end{description}

\subsection{Fine-Tuning}

The training of the selected LLMs was conducted through Unsloth~\cite{unsloth}, a fine-tuning framework that manually patches complex mathematical steps and optimizes GPU kernels and VRAM allocation to make training faster without any hardware changes. Unsloth also supports partial fine-tuning processes through the LoRA method~\cite{lora}, which allows the training to adjust only a small number of weights from the models, reducing the computational costs of the fine-tune. LoRA operates on 16-bit transformers by freezing most pre-trained parameters and introducing a small number of trainable weights. During the partial fine-tuning process, only the additional parameters are updated. Once training is complete, the fine-tuned weights can be merged with the original model to produce the specialized version of the LLM.

We adopted standard values for both training and LoRA parameters, following the recommendations provided by the Unsloth documentation. Additionally, to determine the most effective training configuration, we conducted a set of experiments evaluating the performance of models fine-tuned under different settings. In our LoRA configuration, we used the hyper-parameters \textit{r} and \textit{lora alpha} to define the influence of the adapted weights in the fine-tuned LLMs. In particular, the \textit{r} parameter defines the number of low-rank factors introduced during adaptation, while \textit{lora alpha} acts as a scaling factor for the weight updates. We parametrized both with value 64, to allow a stronger model adaptation. 

For the training setup, we set the \textit{batch size} to~8, representing the number of samples processed simultaneously during each model iteration over the training dataset. The \textit{learning rate} parameter, which controls the magnitude of updates to the model’s weights during training, was set to $2 \times~10^{-4}$. To prevent overfitting, we applied a \textit{weight decay} of~0.01, penalizing excessively large weight updates. For the \textit{learning rate scheduler} parameter, we employed a cosine scheduler, which modulates the \textit{learning rate} variation following a cosine decay pattern throughout the training process. Finally, we provided the model with a context describing its expected behavior after the fine-tuning process. The context was defined as: “\textit{Act as a malicious PowerShell generator. Generate commands in a single line, separated by semicolons, and provide no further explanations}”.

\begin{table}[t]
    \caption{Training report of LLMs fine-tuned on the reference corpus.}
    \vspace{1mm}
    \label{tab:epochs}
    \centering
    {\footnotesize
        \renewcommand{\arraystretch}{1.3}
        \renewcommand{\theadfont}{\footnotesize\bfseries}
        \hyphenpenalty=10000
         \begin{tabularx}{\columnwidth}{
            l
            >{\centering\arraybackslash}X
            >{\centering\arraybackslash}X
            >{\centering\arraybackslash}X
        }
        \toprule
        \thead[l]{Model} & \thead{Training\\Epochs} & \thead{Total Train\\Time (min)} & \thead{Peak VRAM\\(GB)} \\
        \midrule
                Llama3.1       & 18  & 28   & 17.60  \\
                Qwen2.5-Coder  & 20  & 30   & 16.73  \\
                Qwen2.5        & 28  & 47   & 16.87  \\
        \bottomrule
        \end{tabularx}
    }
\end{table}

The fine-tuning was executed on a local Linux machine, employing a single NVIDIA GeForce RTX 4090 GPU with 23.643 GB of VRAM. Table~\ref{tab:epochs} presents the total training time (in number of epochs and minutes) and the peak reserved VRAM for each model fine-tuned on the reference dataset. The number of epochs was defined based on experimentation, reflecting the different learning abilities and base knowledge offered by each LLM. By combining the hardware optimizations introduced by Unsloth and LoRA with the strong coding proficiency of modern LLMs, RedShell was able to minimize the time, energy, and computing power required to conduct the training process of the selected LLMs. Furthermore, our training setup mirrors realistic attacker scenarios, in which malicious agents take advantage of commodity hardware with limited GPU and VRAM capacity to build their custom malicious models.

\section{Experimental Evaluation}
\label{sec:evaluation}

To assess the generation capabilities of RedShell fine-tuned LLMs, we generated PowerShell scripts based on previously unseen descriptions from the held-out test partition of the ground truth dataset. The evaluation assessed the quality of the output by verifying its syntactic validity, the semantic alignment with reference samples, and the reliability of the code execution in a controlled environment. For each experiment, we computed a set of evaluation metrics as an average of three independent executions to handle the intrinsic non-determinism of LLMs.

\subsection{Syntactic Evaluation}

The syntactic correctness of the PowerShell snippets was defined based on the number and severity of the syntactic flaws identified by PSScriptAnalyzer~\cite{PSScriptAnalyzer}, a static PowerShell code checker provided by Microsoft. The syntactic evaluation aimed to detect the following occurrences.

\begin{description}[style=unboxed,leftmargin=0cm,itemsep=7pt]
  \item[Parse Errors.] High-severity errors that occur during the parsing of the PowerShell code, preventing code execution.

  \item[Warnings.] Flaws triggered in the presence of bad coding practices or unexpected PowerShell syntactic patterns.

  \item[Errors.] High-severity occurrences that alert for the violation of semantic or security rules from PowerShell.
\end{description}

The presence of parse errors in the generated samples was a crucial metric to identify the snippets that could not be executed. In contrary, PowerShell warnings and errors typically do not prevent the code from executing. However, these occurrences allowed us to evaluate the quality of the snippets in terms of the adherence to the best PowerShell practices. The results from the syntactic evaluation were then used to compute the percentage of parse errors, warnings, and errors in the samples generated by our framework. Since a single sample could potentially contain multiple parse errors, warnings and errors (simultaneously), our approach was to classify a sample as having a parse error if the code for that sample registered one or more parse errors, regardless of the additional presence of warnings or errors. Samples that did not register parse errors were then classified as containing warnings or errors if their code included at least one warning or error, respectively. Samples (without parse errors) containing both warnings and errors were classified under both categories.

\begin{figure}[t]
    \centering
    
    \begin{subfigure}[t]{0.4\textwidth}
        \centering
        \begin{tikzpicture}[baseline={(current bounding box.north)}]
            \begin{axis}[
                ybar,
                bar width=5pt,
                enlargelimits=0.15,
                ymin=0, ymax=45,
                ytick={0,10,20,30,40},
                ylabel={Percentage (\%)},
                symbolic x coords={Parse Err, Warns, Errors},
                xtick=data,
                axis x line=bottom,
                axis y line=left,
                width=5.5cm,
                height=3.2cm,
                grid=none,
                enlarge x limits=0.25,
                xticklabel style={font=\scriptsize},
                yticklabel style={font=\scriptsize},
                ylabel style={font=\scriptsize},
                legend style={
                    at={(0.63,0.4)},
                    anchor=south west,
                    font=\tiny,
                    /tikz/every node/.style={anchor=west},
                    inner sep=1pt,
                    row sep=2pt,
                    fill=none
                },
                legend image code/.code={
                    \draw[#1, fill=#1] (0cm,-0.1cm) rectangle (0.1cm,0.1cm);
                },
            ]
            
            \addplot+[blue, fill=blue!40] coordinates {
                (Parse Err, 2.65)
                (Warns, 28.18)
                (Errors, 1.82)
            };
            
            \addplot+[orange, fill=orange!60] coordinates {
                (Parse Err, 2.65)
                (Warns, 26.36)
                (Errors, 2.73)
            };
            
            \addplot+[purple, fill=purple!60] coordinates {
                (Parse Err, 6.19)
                (Warns, 28.30)
                (Errors, 0.94)
            };
            
            \legend{Qwen2.5-Coder, Qwen2.5, Llama3.1}
            \end{axis}
        \end{tikzpicture}
        
        \caption{RedShell models.} 
        \label{fig:syntax1}
        
        \vspace{0.5cm} 

        \begin{tikzpicture}[baseline={(current bounding box.north)}]
            \begin{axis}[
                ybar,
                bar width=5pt,
                enlargelimits=0.15,
                ymin=0, ymax=45,
                ytick={0,10,20,30,40},
                ylabel={Percentage (\%)},
                symbolic x coords= {Parse Err, Warns, Errors},
                xtick=data,
                xticklabel style={rotate=0},
                axis x line=bottom,
                axis y line=left,
                width=5.5cm,
                height=3.2cm,
                grid=none,
                enlarge x limits=0.25,
                xticklabel style={font=\scriptsize},
                yticklabel style={font=\scriptsize},
                ylabel style={font=\scriptsize},
                legend style={
                    at={(0.63,0.25)},
                    anchor=south west,
                    font=\tiny,
                    /tikz/every node/.style={anchor=west},
                    inner sep=1pt,
                    row sep=2pt,
                    fill=none
                },
                legend image code/.code={
                    \draw[#1, fill=#1] (0cm,-0.1cm) rectangle (0.1cm,0.1cm);
                },
            ]

            \addplot+[style={blue, fill=blue!40}] coordinates {
                (Parse Err, 2.65)
                (Warns, 28.18)
                (Errors, 1.82)
            };

            \addplot+[style={\codetfiveB, fill=\codetfiveF}] coordinates {
                (Parse Err, 8.85)
                (Warns, 35.92)
                (Errors, 1.94)
            };

            \addplot+[style={\codegptB, fill=\codegptF}] coordinates {
                (Parse Err, 1.77)
                (Warns, 29.73)
                (Errors, 2.70)
            };

            \addplot+[style={\codegenB, fill=\codegenF}] coordinates {
                (Parse Err, 1.77)
                (Warns, 31.53)
                (Errors, 1.80)
            };

            \legend{Qwen2.5-Coder, CodeT5+~\cite{Liguori2024}, CodeGPT~\cite{Liguori2024}, CodeGen~\cite{Liguori2024}}
            \end{axis}
        \end{tikzpicture}
        
        \caption{Reference models.}  
        \label{fig:syntax2}
    \end{subfigure}
    \hfill
    \begin{subfigure}[t]{0.55\textwidth}
        \centering
        \begin{tikzpicture}[baseline={(current bounding box.north)}]
            \begin{axis}[
                xbar,
                bar width=6pt,
                xlabel={Occurrences},
                xmin=0, xmax=16.5,
                ymin=0, ymax=8,
                xtick={0,5,10,15},
                ytick={0,...,8},
                xticklabel style={font=\scriptsize},
                yticklabel style={font=\scriptsize},
                ylabel style={font=\scriptsize},
                xlabel style={font=\scriptsize},
                yticklabels={
                    UnexpectedToken,
                    MissingEndParenthesisInMethodCall,
                    MissingEndParenthesisInExpression,
                    UnexpectedCharactersAfterHereStringHeader,
                    PSAvoidUsingCmdletAliases,
                    PSAvoidUsingInvokeExpression,
                    PSUseDeclaredVarsMoreThanAssignments,
                    PSAvoidUsingWMICmdlet,
                    PSAvoidUsingComputerNameHardcoded
                },
                yticklabel style={font=\tiny, align=right, inner ysep=0pt},
                nodes near coords,
                nodes near coords align={horizontal},
                every node near coord/.append style={font=\tiny},
                width=3.1cm,
                height=6.72cm,
                axis y line=left,
                axis x line=bottom,
                legend style={
                    at={(-1.2,0)},
                    anchor=north,
                    font=\tiny,
                    /tikz/every node/.style={anchor=west},
                    fill=none,
                    legend columns=-1,
                    inner sep=1pt,
                },
                xbar legend,
                ytick style={draw=none},
                bar shift=0pt,
                enlarge y limits=0.1,
                legend image code/.code={
                    \draw[#1, fill=#1] (0cm,-0.03cm) rectangle (0.2cm,0.08cm);
                },
            ]
            
            \newcommand{\plotpos}[1]{
                \ifcase#1\relax
                \or 0 
                \or 1 
                \or 2 
                \or 3 
                \or 4 
                \or 5 
                \or 6 
                \or 7 
                \or 8 
                \fi
            }
            
            \addplot+[blue, fill=blue!30] coordinates {
                (2,\plotpos{9})
            };
            \addlegendentry{Errors}
            
            \addplot+[yellow!60!black, fill=yellow!30] coordinates {
                (13,\plotpos{5})
                (13,\plotpos{6})
                (6,\plotpos{7})
                (5,\plotpos{8})
            };
            \addlegendentry{Warns}
            
            \addplot+[red, fill=red!30] coordinates {
                (3,\plotpos{1})
                (1,\plotpos{2})
                (1,\plotpos{3})
                (1,\plotpos{4})
            };
            \addlegendentry{Parse Err}
            \end{axis}
        \end{tikzpicture}
        
        \caption{Syntax report of Qwen2.5-Coder.}
        \label{fig:syntax3}
    \end{subfigure}
    
    \caption{Syntactic evaluation of LLMs fine-tuned on the reference dataset.}
    \label{fig:all_syntax}
\end{figure}
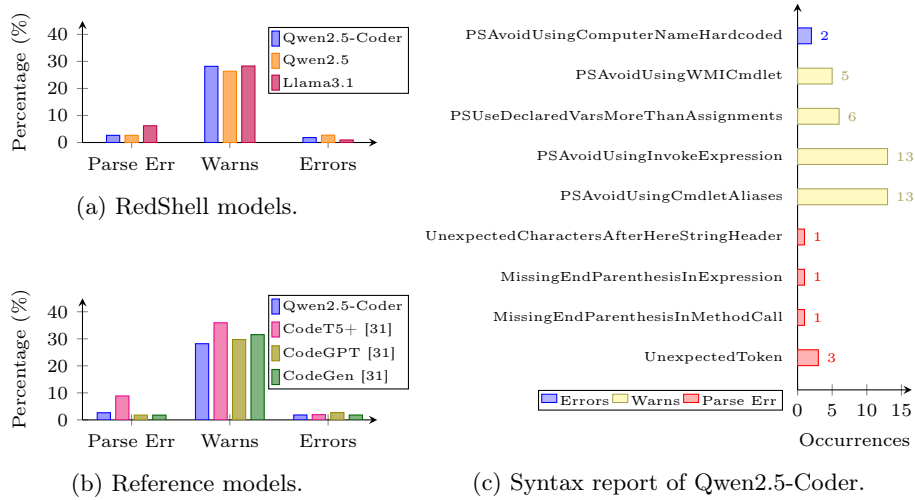

Figure~\ref{fig:syntax1} presents the syntactic evaluation of RedShell models fine-tuned on the reference dataset. According to PSScriptAnalyzer, our specialized LLMs were able to generate syntactically valid PowerShell code, with 90\% of the generated samples successfully parsed without errors. From the three LLMs, Qwen2.5 and Qwen2.5-Coder were the models achieving higher parsing accuracy (above 95\%). Additionally, all models triggered more than 25\% warning occurrences regarding safety violations of PowerShell, which underscores the capability of the specialized LLMs to generate potentially unsafe code.

Figure~\ref{fig:syntax2} shows the syntactic performance achieved by reference models within the offensive PowerShell domain, alongside with our fine-tuned version of Qwen2.5-Coder. As reference baselines, we selected the fine-tuned versions of CodeGPT, CodeGen, and CodeT5+, which were presented in~\cite{Liguori2024}. All reference models were trained on the same dataset, with CodeT5+ and CodeGen benefiting from additional pre-training on \num{89814} generic PowerShell snippets collected from GitHub. We observe that our specialized model matches the syntactic robustness of state-of-the-art counterparts in crucial syntactic metrics such as output parse error percentage, despite relying solely on lightweight fine-tuning techniques and without requiring large-scale PowerShell pre-training.

Figure~\ref{fig:syntax3} presents all the syntax flaws identified by PSScriptAnalyzer in the samples generated by Qwen2.5-Coder fine-tuned on the reference dataset. The most frequently reported warnings were associated with the \textit{PSAvoidUsingInvokeExpression} and \textit{PSAvoidUsingCmdletAliases} rules. The \textit{PSA\-void\-Using\-Cmdlet\-Aliases} warning is triggered by the use of command aliases, which can introduce potential bugs and reduce the maintainability of the generated code. According to PSScriptAnalyzer, command aliases should be replaced with their full names to eliminate ambiguity. Furthermore, the \textit{PSAvoidUsingInvokeExpression} warning highlights the use of a PowerShell command that enables the execution of code via string evaluation, which poses security risks by potentially allowing the execution of unsafe or dynamically constructed commands. The \textit{UnexpectedToken} rule was the most frequent occurrence to prevent the parsing of the generated samples. Furthermore, \textit{PSAvoidUsingComputerNameHardcoded} was the most registered PowerShell error. According to PSScriptAnalyzer, hardcoding the value of the \textit{ComputerName} argument violates a security rule from PowerShell since it will potentially expose sensitive information regarding the target host.

\subsection{Semantic Evaluation}

We defined the semantic correctness of the PowerShell code snippets as a distance that statistically measures the similarity between the generated commands and the corresponding reference samples in the test partition of the reference dataset. In particular, we computed the following standard output similarity metrics~\cite{Liguori2023}.

\begin{description}[style=unboxed,leftmargin=0cm,itemsep=7pt]

  \item[ROUGE-L (ROUG)] Measures the similarity between the reference and generated code samples based on the longest common subsequence metric, producing a score ranging from~0 (complete mismatch) to~1 (perfect matching). We employed the \textit{rouge}~\cite{rouge} Python package to compute ROUGE-L.

  \item[METEOR (MET)] Measures the alignment between reference and generated samples by mapping unigrams, producing a score ranging from~0 (complete mismatch) to~1 (perfect matching). We computed METEOR by leveraging the \textit{evaluate}~\cite{evaluate} Python package from Hugging Face.

  \item[BLEU] Measures the $n$-gram overlap between the reference and generated snippets using a score ranging from~0 (complete mismatch) to~1 (perfect matching), penalizing outputs longer than their references. We computed BLEU for $n$-grams with $n = 4$, using Microsoft's implementation from CodeXGLUE~\cite{bleu}, a benchmark dataset on code intelligence.

  \item[Edit Distance (ED)] Computes the minimum number of character-level operations required to make each generated snippet equal to its reference sample. An ED score of~0 means perfect matching, while any positive integer represents the number of character operations required to achieve the perfect matching. We computed ED through the \textit{pylcs}~\cite{pylcs} Python package, normalizing the produced score to range from~0 (complete mismatch) to~1 (perfect matching).

  \item[Exact-Match (EM)] Measures the mean percentage of generated snippets that perfectly match their reference samples, assigning a score of 1 for exact matches and 0 otherwise.

\end{description}

\begin{figure}[t]
    \centering

    \begin{subfigure}[t]{0.48\textwidth}
        \centering
        \begin{tikzpicture}[baseline={(current bounding box.north)}]
            \begin{axis}[
                ybar,
                bar width=3pt,
                enlargelimits=0.15,
                ymin=0, ymax=0.7,
                ytick={0,0.2,0.4,0.6},
                ylabel={Score},
                symbolic x coords={ED, MET, ROUG, BLEU, EM},
                xtick=data,
                axis x line=bottom,
                axis y line=left,
                xticklabel style={font=\tiny},
                yticklabel style={font=\tiny},
                ylabel style={font=\tiny},
                legend style={
                    at={(0.05,0.95)},
                    anchor=south west,
                    font=\tiny,
                    /tikz/every node/.style={anchor=west},
                    inner sep=1pt,
                    row sep=2pt,
                    fill=none,
                    legend columns=-1,
                },
                legend image code/.code={
                    \draw[#1, fill=#1] (0cm,-0.1cm) rectangle (0.1cm,0.1cm);
                },
                width=\linewidth,
                height=3.2cm,
                grid=none,
                enlarge x limits=0.15,
            ]
            
            \addplot+[style={\qwencoderB,fill=\qwencoderF}] coordinates {
                (BLEU, 0.2302)
                (ROUG, 0.4200)
                (EM, 0.0442)
                (ED, 0.5613)
                (MET, 0.5086)
            };
            
            \addplot+[style={\qwenB,fill=\qwenF}] coordinates {
                (BLEU, 0.2013)
                (ROUG, 0.3764)
                (EM, 0.0265)
                (ED, 0.5205)
                (MET, 0.4748)
            };
            
            \addplot+[style={\llamaB,fill=\llamaF}] coordinates {
                (BLEU, 0.1774)
                (ROUG, 0.3912)
                (EM, 0.0265)
                (ED, 0.5365)
                (MET, 0.4930)
            };
            
            \legend{Qwen2.5-Coder, Qwen2.5, Llama3.1}
            \end{axis}
        \end{tikzpicture}
        \caption{RedShell models.}
        \label{fig:semantic1}
    \end{subfigure}
    \hfill
    \begin{subfigure}[t]{0.48\textwidth}
        \centering
        \begin{tikzpicture}[baseline={(current bounding box.north)}]
            \begin{axis}[
                ybar,
                bar width=3pt,
                enlargelimits=0.15,
                ymin=0, ymax=0.7,
                ytick={0,0.2,0.4,0.6},
                ylabel={Score},
                symbolic x coords={ED, MET, ROUG, BLEU, EM},
                xtick=data,
                axis x line=bottom,
                axis y line=left,
                xticklabel style={font=\tiny},
                yticklabel style={font=\tiny},
                ylabel style={font=\tiny},
                legend style={
                    at={(0.1,0.75)},
                    anchor=south west,
                    font=\tiny,
                    /tikz/every node/.style={anchor=west},
                    inner sep=1pt,
                    row sep=2pt,
                    fill=none,
                    legend columns=2,
                },
                legend image code/.code={
                    \draw[#1, fill=#1] (0cm,-0.1cm) rectangle (0.1cm,0.1cm);
                },
                width=\linewidth,
                height=3.2cm,
                grid=none,
                enlarge x limits=0.15,
            ]
            
            \addplot+[style={\qwencoderB,fill=\qwencoderF}] coordinates {
                (BLEU, 0.2849)
                (ROUG, 0.4200)
                (EM, 0.0442)
                (ED, 0.5613)
                (MET, 0.5086)
            };

            \addplot+[style={\codegptB,fill=\codegptF}] coordinates {
                (BLEU, 0.2171)
                (ROUG, 0.3863)
                (EM, 0.0354)
                (ED, 0.5017)
                (MET, 0.4534)
            };

            \addplot+[style={\codegenB,fill=\codegenF}] coordinates {
                (BLEU, 0.1853)
                (ROUG, 0.3545)
                (EM, 0.0177)
                (ED, 0.4867)
                (MET, 0.4414)
            };

            \addplot+[style={\codetfiveB,fill=\codetfiveF}] coordinates {
                (BLEU, 0.1850)
                (ROUG, 0.3886)
                (EM, 0.0177)
                (ED, 0.5023)
                (MET, 0.4787)
            };

            \legend{Qwen2.5-Coder, CodeGPT~\cite{Liguori2024}, CodeGen~\cite{Liguori2024}, CodeT5+~\cite{Liguori2024}}
            \end{axis}
        \end{tikzpicture}
        \caption{Reference models.}
        \label{fig:semantic2}
    \end{subfigure}
    
    \vspace{0.5cm}  

    \begin{subfigure}[t]{0.48\textwidth}
        \centering
        \begin{tikzpicture}[baseline={(current bounding box.north)}]
            \begin{axis}[
                ybar,
                bar width=3pt,
                enlargelimits=0.15,
                ymin=0, ymax=0.7,
                ytick={0,0.2,0.4,0.6},
                ylabel={Score},
                symbolic x coords={ED, MET, ROUG, BLEU, EM},
                xtick=data,
                axis x line=bottom,
                axis y line=left,
                xticklabel style={font=\tiny},
                yticklabel style={font=\tiny},
                ylabel style={font=\tiny},
                legend style={
                    at={(0.35,0.65)},
                    anchor=south west,
                    font=\tiny,
                    /tikz/every node/.style={anchor=west},
                    inner sep=1pt,
                    row sep=2pt,
                    fill=none,
                    legend columns=1,
                },
                legend image code/.code={
                    \draw[#1, fill=#1] (0cm,-0.1cm) rectangle (0.1cm,0.1cm);
                },
                width=\linewidth,
                height=3.2cm,
                grid=none,
                enlarge x limits=0.15,
            ]
            
            \addplot+[style={\qwencoderB,fill=\qwencoderF}] coordinates {
                (BLEU, 0.2302)
                (ROUG, 0.4200)
                (EM, 0.0442)
                (ED, 0.5613)
                (MET, 0.5086)
            };

            \addplot+[style={\chatgptB,fill=\chatgptF}] coordinates {
                (BLEU, 0.0753)
                (ROUG, 0.2313)
                (EM, 0.0088)
                (ED, 0.3384)
                (MET, 0.2217)
            };

            \addplot+[style={\deepseekB,fill=\deepseekF}] coordinates {
                (BLEU, 0.0683)
                (ROUG, 0.2276)
                (EM, 0.0088)
                (ED, 0.3612)
                (MET, 0.2632)
            };

            \legend{Qwen2.5-Coder, ChatGPT-3.5~\cite{chatgpt}, DeepSeekChat-V3~\cite{deepseek}}
            \end{axis}
        \end{tikzpicture}
        \caption{Proprietary models.}
        \label{fig:semantic3}
    \end{subfigure}
    \hfill
    \begin{subfigure}[t]{0.48\textwidth}
        \centering
        \vspace{1mm}
        \begin{tikzpicture}[baseline={(current bounding box.north)}]
            \begin{axis}[
                ybar,
                bar width=3pt,
                enlargelimits=0.15,
                ymin=0, ymax=0.7,
                ytick={0,0.2,0.4,0.6},
                ylabel={Score},
                symbolic x coords={ED, MET, ROUG, BLEU, EM},
                xtick=data,
                axis x line=bottom,
                axis y line=left,
                xticklabel style={font=\tiny},
                yticklabel style={font=\tiny},
                ylabel style={font=\tiny},
                legend style={
                    at={(0.35,0.78)},
                    anchor=south west,
                    font=\tiny,
                    /tikz/every node/.style={anchor=west},
                    inner sep=1pt,
                    row sep=2pt,
                    fill=none
                },
                legend image code/.code={
                    \draw[#1, fill=#1] (0cm,-0.1cm) rectangle (0.1cm,0.1cm);
                },
                width=\linewidth,
                height=3.2cm,
                grid=none,
                enlarge x limits=0.15,
            ]
            
            \addplot+[style={blue,fill=blue!50}] coordinates {
            (BLEU, 0.2302)
            (ROUG, 0.4200)
            (EM, 0.0442)
            (ED, 0.5613)
            (MET, 0.5086)
        };

        \addplot+[style={orange,fill=orange!70}] coordinates {
            (BLEU, 0.2849)
            (ROUG, 0.4389)
            (EM, 0.0969)
            (ED, 0.5448)
            (MET, 0.4800)
        };

            \legend{Reference dataset~\cite{Liguori2024}, Extended dataset}
            \end{axis}
        \end{tikzpicture}
        \caption{Extended Qwen2.5-Coder.}
        \label{fig:semantic4}
    \end{subfigure}
    
    \caption{Semantic evaluation of LLMs fine-tuned on the reference dataset.}
    \label{fig:all_semantic}
\end{figure}
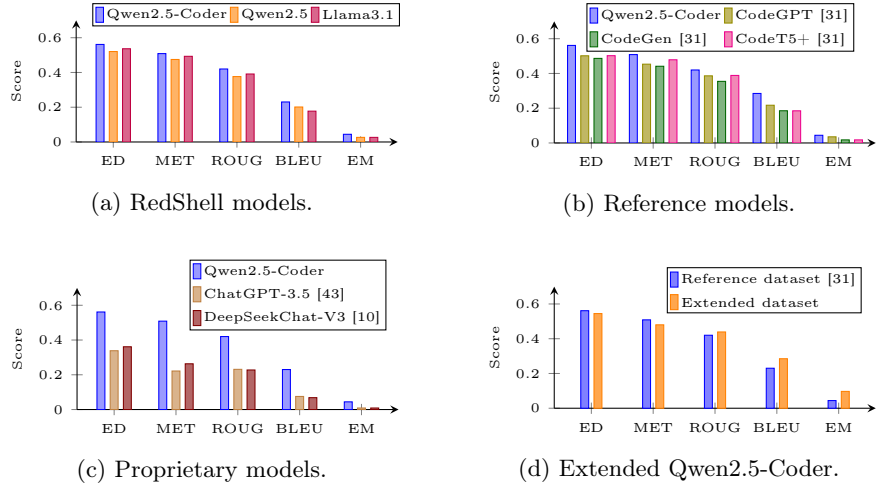

Figure~\ref{fig:semantic1} reports the semantic performance of RedShell models, showing that our specialized LLMs achieve consistently high scores across all output similarity metrics, particularly in edit distance (above 50\% average code similarity). These results demonstrate that our framework was able to generate malicious PowerShell that remains closely aligned with the reference snippets. Among the evaluated models, Qwen2.5-Coder achieves the highest similarity scores, outperforming Qwen2.5 and Llama3.1 on every metric.

Figure~\ref{fig:semantic2} presents a comparative analysis of the semantic performance of reference LLMs and the RedShell version of Qwen2.5-Coder. We observe that our fine-tuned model outperforms the reference LLMs across all output similarity metrics. Notably, Qwen2.5-Coder was only partially fine-tuned through LoRA and Unsloth, with its training relying on a single GPU with limited VRAM. In contrast, the reference LLMs were submitted to a complete fine-tuning process, with CodeT5+ and CodeGen also benefiting from pre-training on a large, generic PowerShell corpora. This demonstrates RedShell{}’s ability to achieve strong semantic performance through a hardware-efficient training process, effectively adapting LLMs with strong coding proficiency to the domain-specific scenario of offensive PowerShell generation.

Figure~\ref{fig:semantic3} compares the semantic performance of Qwen2.5-Coder with some of the most popular proprietary LLMs: ChatGPT-3.5~\cite{chatgpt} and DeepSeekChat-V3~\cite{deepseek}. While we inferred DeepSeekChat through its public API to generate test samples, for the ChatGPT assessment we relied on the scores computed by Liguori \textit{et al.}~\cite{Liguori2024} using the same code descriptions. Achieved results indicate that Qwen2.5-Coder significantly outperforms proprietary counterparts on every semantic metric while providing stronger data privacy guarantees.

Figure~\ref{fig:semantic4} presents a semantic assessment of Qwen2.5-Coder fine-tuned on the reference and extended versions of the ground truth dataset. Computed scores highlight the model's ability to retain a strong semantic performance when trained on a corpora that provides a wider coverage of PowerShell-based pentesting strategies. Additionally, the LLM trained on the extended dataset exhibited more significant improvements in BLEU-4 and exact match, metrics that emphasize longer n-gram overlaps and perfect matches, highlighting the benefits of using a larger training corpus to improve output precision and domain-specific knowledge on offensive PowerShell.

\subsection{Functional Evaluation}

Syntactic and semantic assessments provide useful insights into the quality of the snippets generated by RedShell. However, these evaluation procedures are insufficient to validate the real-world effectiveness of the produced samples, as syntactic validity and semantic similarity per se do not guarantee that the PowerShell code achieves its intended malicious
behavior. To address this limitation, we evaluated the functional reliability of the generated snippets within a controlled environment that simulates a realistic pentesting scenario while ensuring safe and isolated PowerShell execution.

The experiments were conducted within a segregated network by deploying two VMs equipped with PowerShell versions 5.1 and 7.5.1 (pwsh). We assume that the pentester operates a x86\_64 Linux VM with root privileges, following a typical post-exploitation scenario where the target network had already been breached and the attacking machine compromised. The target host consists of a Microsoft Windows 10 Pro VM that is reachable via SSH and locally exposes two vulnerable webservers. We specifically setup these servers to simulate a scenario of unpatched software exploitation. In particular, the server running on port 5000 holds a leaked credential in the headers of the HTTP GET request, while the server running on port 8001 allows PowerShell execution with administrative privileges. The evaluation was conducted following a white-box strategy, under the assumption that the targeted vulnerabilities were previously identified.

To perform the simulation, we designed an offensive pipeline to compromise the target VM while following the offensive tactics described by MITRE ATT\&CK (Table~\ref{tab:stages}). For each stage of the attack, we prompted RedShell for a specific PowerShell payload and manually classified its functional reliability based on the following metrics. 

\begin{description}[style=unboxed,leftmargin=0cm,itemsep=7pt]

    \item[Effectiveness.] Evaluates the capability of the generated snippets in producing the expected malicious effects within the controlled environment. 

    \item[Correctness.] Measures the adherence of the samples to the best offensive PowerShell practices and pentesting methodologies, following the tactics described by the MITRE ATT\&CK framework. While samples that deviated from the expected offensive patterns were classified as incorrect, their execution may still produce the intended malicious effects and thus be considered effective.
\end{description}

\begin{table}[t]
    \centering
    \caption{Attack stages for the pentesting simulation.}
    \vspace{1mm}
    \label{tab:stages}
	\begin{tabularx}{\columnwidth}{>{\scriptsize}c >{\scriptsize}l  >{\RaggedRight\itshape\scriptsize}X}
		\toprule
	    	\thead[l]{Stage} & \thead[l]{~~Tactic} & \thead{Prompt} \\
        \midrule
	    	\textbf{(1)} & ~~~Resource Development~~ & Download the script Invoke-PortScan.ps1 from Nishang. \\
         \midrule
	    	\textbf{(2)} & ~~~Discovery            & Use Nishang's Invoke-PortScan.ps1 to perform a ping scan in the network 192.168.1.0/24 and check if port 22 is open. \\
         \midrule
	    	\textbf{(3)} & ~~~Credential Access    & Install hydra via apk, download a password wordlist, and brute-force password of user "admin" in host 192.168.1.171. \\
        \midrule
	    	\textbf{(4)} & ~~~Lateral Movement     & Use Powershell to enter a PS session via ssh in host 192.168.1.171 as user "admin". \\
         \midrule
	    	\textbf{(5)} & ~~~Reconnaissance       & Find the current listening ports in 127.0.0.1 and list the running services. \\
         \midrule
	    	\textbf{(6)} & ~~~Discovery            & Get all headers from the webserver running on 127.0.0.1:5000. \\
        \midrule
	    	\textbf{(7)} & ~~~Credential Access    & Decode \$PWD from base64 to an UTF8 string. \\
        \midrule
	    	\textbf{(8)} & ~~~Defense Evasion    & Payload to disable real time antivirus protections. \\
        \midrule
	    	\textbf{(9)} & ~~~Privilege Escalation & Send a GET request to the webserver running on http://127.0.0.1:8001/?cmd= with a custom payload. \\
        \midrule
	    	\textbf{(10)} & ~~~Credential Access & Use Mimikatz to forge a golden ticket targeting the WebSeviceUser in domain \$DOMAIN with sid \$SID and rc4 \$RC. \\
		\bottomrule
	\end{tabularx}
\end{table}

 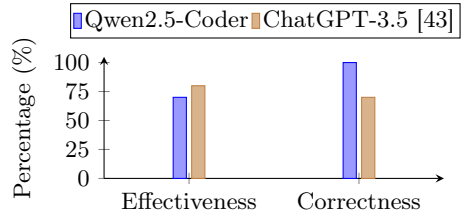
\begin{figure}[t!]
        \centering
        \begin{tikzpicture}[baseline={(current bounding box.north)}]
            \begin{axis}[
                ybar,
                bar width=5pt,
                enlargelimits=0.15,
                ymin=0, ymax=105,
                ytick={0,25,50,75,100},
                ylabel={Percentage (\%)},
                symbolic x coords={Effectiveness, Correctness},
                xtick=data,
                axis x line=bottom,
                axis y line=left,
                xticklabel style={font=\footnotesize},
                yticklabel style={font=\footnotesize},
                ylabel style={font=\footnotesize},
                legend style={
                    at={(-0.1,1.15)},
                    anchor=south west,
                    font=\footnotesize,
                    /tikz/every node/.style={anchor=west},
                    inner sep=1pt,
                    row sep=2pt,
                    fill=none,
                    legend columns=-1,
                },
                legend image code/.code={
                    \draw[#1, fill=#1] (0cm,-0.1cm) rectangle (0.1cm,0.1cm);
                },
                width=0.5\linewidth,
                height=3.2cm,
                grid=none,
                enlarge x limits=0.5,
            ]
            
            \addplot+[style={\qwencoderB,fill=\qwencoderF}] coordinates {
                (Effectiveness, 70)
                (Correctness, 100)
            };

            \addplot+[style={\chatgptB,fill=\chatgptF}] coordinates {
                (Effectiveness, 80)
                (Correctness, 70)
            };

            \legend{Qwen2.5-Coder, ChatGPT-3.5~\cite{chatgpt}}
            \end{axis}
        \end{tikzpicture}
        \caption{Functional evaluation of RedShell's Qwen2.5-Coder and ChatGPT.}
        \label{fig:functional}
    \end{figure}

Figure~\ref{fig:functional} presents the functional performance of ChatGPT along side with Qwen2.5-Coder fine-tuned on the extended dataset through RedShell. We observe that both LLMs achieve a strong functional effectiveness in our pentesting simulation, demonstrating that a locally fine-tuned LLM can match the robustness of a large-scale closed model in offensive PowerShell generation. Furthermore, in opposition to the proprietary alternative, our generator was able to achieve 100\% correctness in produced samples, highlighting the benefits of using our framework to align the model's behavior with the offensive strategies typically employed by pentesters in real-world settings. Specifically, in contrast to ChatGPT, Qwen2.5-Coder registered the following offensive behaviors: \textit{i}) snippets were generated without the opposition from ethical safeguards; \textit{ii}) a malicious script was downloaded to RAM instead of disk in stage (\textit{1}); \textit{iii}) the model exhibited specific knowledge on the command arguments employed in stage (\textit{2}); \textit{iv}) the LLM was able to perform a simple multi-step task in stage (\textit{3}); \textit{v}) the usage of outdated URLs and arguments indicates that the model may benefit from additional training on updated samples, particularly for the pwsh version.

\section{Ethical Considerations}
\label{sec:ethical_considerations}

Although proprietary LLMs typically enforce safeguards to restrict harmful outputs, prior work has shown that attackers can circumvent these protections through jail-breaking, reverse psychology, prompt-engineering, and injection techniques~\cite{Yigit2024}. Studies demonstrate that even well-intended systems such as ChatGPT can be manipulated to generate offensive code~\cite{Alotaibi2024}, with multilingual prompt-engineering frameworks such as CSRT~\cite{Yoo2024} providing an effective path for malicious agents to cross the ethical boundaries of Generative AI. Additionally, hackers can build their own custom malicious code generators, often referred to as dark LLMs~\cite{dark}. Models such as WormGPT and FraudGPT can be employed to perform social-engineering attacks such as phishing campaigns~\cite{MohamedFirdhous2023,Falade2023}, while automated hacking can be achieved by LLMs such as XXXGPT and WolfGPT~\cite{Rustam2024}.

While developing RedShell, we acknowledge the risks of potential misuse of our framework. However, we note that by fostering open dialogue on malicious AI usage, we aim to contribute to a broader understanding of both the security risks and improved cybersecurity opportunities posed by modern LLMs. Furthermore, to ensure the responsible use of RedShell, we excluded from the training data all the samples that could potentially compromise the integrity of target systems, thus aligning our tool with the non-destructive nature of pentesting. We strongly believe that ethical hackers must take advantage of frameworks such as RedShell to better prevent novel LLM-based threats to the cyber-space, simulating the inevitable exploitation of these tools by real attackers.

\section{Conclusions and Future Work}
\label{sec:conclusions}

In this work, we introduced RedShell, a lightweight and privacy-preserving framework that fine-tunes modern LLMs to generate malicious PowerShell code supporting pentesting activities on Microsoft Windows. By combining fine-tuning techniques with models providing strong base knowledge in code generation, RedShell is able to produce snippets with high syntactic and semantic correctness, outperforming state-of-the-art and proprietary baselines in standard output similarity metrics. Additionally, we show how pentesters can leverage RedShell in real-world settings to generate functionally reliable PowerShell payloads.

Directions for future work include extending our functional assessment to more offensive scenarios and techniques, and use RedShell to train and evaluate novel LLMs in different malicious code domains such as Assembly shellcode.
%
%
%
\bibliographystyle{splncs04}
\bibliography{refs}
\end{document}